\documentclass[usenatbib]{mn2e}
\usepackage{graphicx}
\usepackage{float}
\usepackage{fixltx2e}
\usepackage[T1]{fontenc} 
\usepackage{aecompl}
\title{HD314884: A Slowly Pulsating B star in a Close Binary}
\author[Christopher B. Johnson et al.]{Christopher B. Johnson$^{1}$\thanks{E-mail: cjoh285@lsu.edu (CBJ)}, R. I. Hynes$^{1}$, T. Maccarone$^{2,7}$, C. T. Britt$^{2}$, H. Davis III$^{1}$,
\newauthor
P. G. Jonker$^{3,4,5}$, M. A. P. Torres$^{3}$, D. Steeghs$^{6}$, S. Greiss$^{6}$, G. Nelemans$^{5}$\\
$^{1}$Department of Physics and Astronomy, Louisiana State University, Baton Rouge, LA 70803-4001 USA\\
$^{2}$Department of Physics, Texas Tech University, Box 41051, Lubbock, TX 79409-1051, USA\\
$^{3}$SRON, Netherlands Institute for Space Research, Sorbonnelaan 2, 3584 CA, Utrecht, The Netherlands\\
$^{4}$Harvard-Smithsonian Center for Astrophysics, 60 Garden Street, Cambridge, MA, 02138, USA\\
$^{5}$Department of Astrophysics/ IMAPP, Radbound University Nijmegen, Heyendaalseweg 135, 6525 AJ, Nijmegen, The Netherlands\\
$^{6}$Department of Physics, University of Warwick, Coventry CV4 7AL, UK\\
$^{7}$Physics and Astronomy, University of Southampton, Highfield, Southampton, SO17 1BJ, UK\\}
\begin{document}

\date{Accepted; Received; in original form}

\pagerange{\pageref{firstpage}--\pageref{lastpage}} \pubyear{2013}

\maketitle

\label{firstpage}

\begin{abstract}We present the results of a spectroscopic and photometric analysis of HD314884, a slowly pulsating B star (SPB) in a binary system with detected soft X-ray emission. We spectrally classify the B star as a B5V--B6V star with T$_{eff}$ = 15,490 $\pm$ 310 K, log $g$ = 3.75 $\pm$ 0.25 dex, and a photometric period of P$_{0}$ = 0.889521(12) days. A spectroscopic period search reveals an orbital period for the system of P$_{orb}$ = 1.3654(11) days. The discrepancy in the two periods  and the identification of a second and third distinct frequency in the photometric fourier transform at P$_1$ = 3.1347(56) and P$_2$ = 1.517(28) days provides evidence that HD314884 is a slowly pulsating B star (SPB) with at least three oscillation frequencies. These frequencies appear to originate from higher-order, non-linear tidal pulsations. Using the dynamical parameters obtained from the radial velocity curve, we find the most probable companion mass to be M$_1$ = $\sim$0.8 M$_{\sun}$ assuming a typical mass for the B star and most probable inclination. We conclude that the X-ray source companion to HD314884 is most likely a coronally active G-type star or a white dwarf (WD), with no apparent emission lines in the optical spectrum. The mass probability distribution of the companion star mass spans 0.6--2.3 M$_{\sun}$ at 99$\%$ confidence which allows the possibility of a neutron star companion. The X-ray source is unlikely to be a black hole unless it is of a very low mass or low binary inclination.
\end{abstract}


\section{Introduction} Slowly pulsating B stars (SPB)  are typically mid- to late-type B stars that show photometric periods from 0.5--5 days and masses ranging from 4.0--7.0 M$_{\sun}$ \citep{wel91}. The effective temperature range of known SPBs is between 10,000--20,000 K with dense frequency spectra showing low-amplitude multi-periodicity. The pulsations are high order, low degree, gravity mode (g-mode) pulsations that are driven by the $\kappa$-mechanism of the Fe opacity bump at T $\approx$ 1.5 x 10$^5$K \citep{dzi93}. Their existence and close relation to the $\delta$ Sct stars provides an excellent test bed for probing the deepest layers of the stellar atmosphere through the use of asteroseismology.\\
\indent Early type stars have been known to be X-ray sources for quite some time now. The existence of X-ray emission from massive O and B stars was predicted decades ago by \citet{cas79} and was serendipitously discovered at the same time during early observations using the Einstein satellite. The X-ray properties of O and B stars were constrained more accurately by \citet{ber97} using the ROSAT All-Sky Survey (RASS). Equipped with 237 detections, \citet{ber97} confirmed the decline in the detection rate toward later spectral types (all stars of spectral type O7 or earlier were detected as X-ray sources, while at most 10 \% of B3--B9 stars were detected). This fact, and a higher incidence of variability and binarity among the later spectral types, led to the conclusion that low-mass companions could be responsible for the X-ray emission of late-B stars. This can be interpreted in one of two ways. (1) A matter stream exists between the B star and the companion due to Roche lobe overflow (RLOF). The stripped matter impacts the accretion disc causing X-ray emission at the impact point. (2) The low mass companion is coronally active, which is to be expected since such companions must be very young and have not had enough time to slow their rotational speed yet.\\ 
\indent We report the discovery of a new SPB star found in the Galactic Bulge Survey (GBS). The Galactic Bulge Survey \citep{jon11} is a wide and shallow X-ray survey of two strips both above and below the Galactic bulge aimed at detecting such X-ray emission. The campaign is designed to look for X-ray binaries (XRB), specifically, low-mass X-ray binaries (LMXBs) harboring black holes (BHs) or neutron stars (NS) to allow for mass measurements and the comparison of the number counts and orbital period distribution of these sources with the predictions of population synthesis modelling. Along with these compact objects, a number of GBS sources were identified with optically bright stars which include, but are not limited to, late B and A stars \citep{hyn12}.\\
\indent The GBS X-ray source CXOGBS J175637.0-271145 = CX514 ($V$ = 10.04 mag) has coordinates that align with HD314884, initially classified as a B9 star \citep{nes95}. CX514 was found to have an X-ray luminosity of at least 1.3$\times$10$^{30}$ erg s$^{-1}$ and log($\frac{F_X}{F_{bol}}$) = -5.4  \citep{hyn12}, neither of which is consistent with a single late-type B star.  We infer that this is either a chance alignment with HD314884 or that the X-ray source is the binary companion to the B star. The 95\% confidence radius of uncertainty in the Chandra position is 3.4". The optical counterpart to CX514 (HD314884) has also been identified as a new All Sky Automated Survey (ASAS) Catalogue \citep{poj02} variable source ASAS ID\# 175637-2711.8 suggesting it may be an unusual source.\\
\indent In this paper, we analyze the binary system composed of HD314884 and its companion. We present phase-resolved optical spectroscopy and photometry of the optical counterpart drawn from several sources. Orbital parameters derived for the system are then used to constrain the mass of the X-ray source. We begin with a short introduction followed by discussing the observations, both photometric and spectroscopic, in section 2. Section 3 discusses spectral classification and features of HD314884. Section 4 describes the period analysis, whereas the radial velocity curve parameters and mass function are determined in Section 5. Finally, the nature of the optical counterpart and the X-ray source is discussed in Section 6 followed by the conclusion.

\section[]{Observations and Data Reduction}
\subsection{ASAS Archival Data} We began by collecting all the available $V$-band data from the ASAS All Star Catalogue of Photometry. ASAS monitored HD314884 from 2001 until 2009 and is found in the ASAS-3 photometric $V$ band catalogue with data spanning MJD 51949.87862--55138.50718. We performed our final analysis using only the grade A or B observations for the smallest aperture which included 720 data points. This is the same data sample used in \cite{hyn12}.      	
\subsection{0.9 meter SMARTS Optical Photometry} 
We observed HD314884 on 2012 June 1 until 2012 June 8 using the 0.9 meter SMARTS Consortium telescope at Cerro Tololo Inter-American Observatory (CTIO) with an aim at ruling out one day aliases inherent in the ASAS data. We collected 22 observations in the SDSS $r'$ filter with an average exposure time of 60 sec. Sky flats were obtained in the beginning of the night in each of the filters with varying exposure times. Biases and dome flats were obtained at the end of every night. Standard data reduction techniques were used (bias subtraction, flat-fielding, etc.) with the aid of IRAF\footnote[1]{IRAF is distributed by the National Optical Astronomy Observatory, which is operated by the Association of Universities for Research in Astronomy (AURA) under cooperative agreement with the National Science Foundation.} packages and tools.\\  
\subsection{Optical Spectroscopy} Optical spectroscopy was obtained with the 2.0 m Liverpool Telescope at the Observatorio del Roque de Los Muchachos in La Palma, Spain. Spectra were collected using the Fibre-fed Robotic Dual-beam Optical Spectrograph, or FRODOspec instrument$^{2}$. Table~1 shows the spectroscopic data collection for HD314884.  
\begin{table}
\caption{Table of Spectroscopic Data Collection}
\centering
\begin{tabular}{c c c c}
\hline \hline
Date Observed&UT Start&MJD&Exp. Time (sec)\\
\hline
2012 August 01&20:36:15.442& 56140.858512&    300\\
2012 August 01&20:41:42.834& 56140.862301&    300\\
2012 August 03&23:00:14.400& 56142.958500&    300\\
2012 August 03&23:05:49.381& 56142.962377&    300\\
2012 August 04&22:36:45.401& 56143.942192&    300\\
2012 August 04&22:42:12.978& 56143.945984&    300\\
2012 August 05&23:47:56.222& 56144.991623&    300\\
2012 August 05&23:53:29.390& 56144.995479&    300\\
2012 August 06&22:16:40.483& 56145.928246&    300\\
2012 August 06&22:22:08.115& 56145.932038&    300\\
2012 August 08&00:01:13.356& 56147.000849&    300\\
2012 August 08&00:06:47.980& 56147.004722&    300\\
2012 August 08&22:58:37.248& 56147.957376&    300\\
2012 August 08&23:04:04.668& 56147.961165&    300\\
2012 August 09&22:28:32.045& 56148.936482&    300\\
2012 August 09&22:33:59.714& 56148.940274&    300\\
2012 August 10&21:23:01.776& 56149.890993&    300\\
2012 August 10&21:28:29.879& 56149.894790&    300\\
2012 August 13&21:35:03.811& 56152.899350&    300\\
2012 August 13&21:40:32.820& 56152.903158&    300\\
2012 August 14&22:47:46.790& 56153.949847&    300\\
2012 August 14&22:53:14.352& 56153.953638&    300\\
\hline
\end{tabular}\\
\end{table}
CX514 was observed with both the blue-high grating  with a wavelength range of 3800--5150\,\AA\ and the red-high grating with a wavelength range of 5700-8050 \AA. The dispersion in the blue range is 0.35 \AA/pixel with a resolution of 0.8 \AA\ at the central wavelength of 4495.6 \AA. The dispersion inherent in the red range is 0.58 \AA/pixel with a resolution of 1.3 \AA\ at a central wavelength of 6827 \AA. Exposure times on both sets of observations were 300 seconds each. The spectra were reduced by using two sequentially invoked FRODOspec pipelines. The first pipeline, known as the L1, is a CCD processing pipeline which performs bias subtraction, overscan trimming and CCD flat fielding. The second pipeline, known as the L2, performs the spectroscopic data extraction. The pipeline reduced data have a zero-point magnitude of 15.1 and 14.2 in the AB magnitude system for the red and blue gratings, respectively. 
The wavelength calibrated spectra have a final accuracy of 0.08 \AA. The spectra were then normalised with standard IRAF tools and packages by dividing by a spline fit to the continuum of order 10. The FRODOspec pipeline and instrument specifications are described in detail online\footnote[2]{http://telescope.livjm.ac.uk/Info/TelInst/Inst/FRODOspec}.

  \begin{figure}
	\centering
	\includegraphics[width=3in]{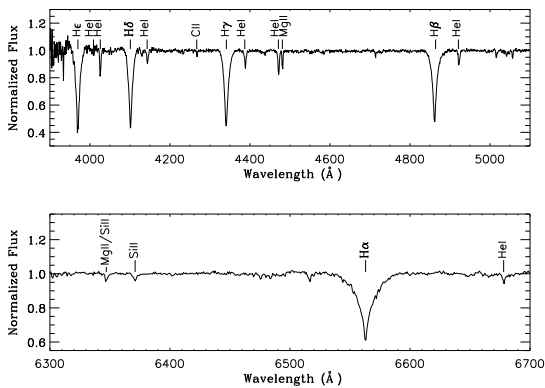}
	\centering
	\caption{Normalised spectra of HD314884 with the top panel showing the blue end and the bottom showing the red end of the spectrum. Prominent spectral lines are marked.}
	\centering
	\label{fig:}
\end{figure} 

\section{Spectral Classification and Features} We begin the spectral classification by comparing the spectral range 3900--5100 {\AA} of the counterpart to several known MK standards of spectral type B. By visual inspection of Fig.~1, we can see the weak presence of several HeI lines along with the CII and MgII lines. The most notable lines would be the Balmer lines with the presence of HeI lines of $\lambda$4026, $\lambda$4143, $\lambda$4387, and $\lambda$4471. Several helium and metal lines can be used to aid in classifying the spectrum of the counterpart. The HeI $\lambda$4009 is prominent down to spectral type B3, then starts to get weaker at B5 before disappearing at B8. This feature is very weak in the HD314884 spectrum. The ratio of HeI $\lambda$4471/MgII $\lambda$4481 is also a useful indicator used to distinguish between mid- to late-type B stars. The neutral HeI $\lambda$4471 line weakens and disappears as the MgII $\lambda$4481 strengthens when approaching lower temperatures \citep{gra2009} in late-type B stars.\\
\indent We measured the HeI $\lambda$4471/MgII $\lambda$4481 ratio to be 1.42 $\pm$ 0.02 for all 22 spectra. We rule out spectral classes B8V-B9V based on the HeI $\lambda$4471/MgII $\lambda$4481 ratio $<$ 1.00 due to the stronger MgII $\lambda$4481 as the T$_{eff}$ approaches cooler temperatures. We then measured the HeI $\lambda$4471/MgII $\lambda$4481 ratio for several MK standards of types B0V--B9V as a comparison. On average, the B0V--B4V spectra had a HeI $\lambda$4471/MgII $\lambda$4481 ratio $>$ 1.55 $\pm$ 0.02. We find that these indicators, along with the weakness of CII $\lambda$4267, point to a B5V--B6V classification for the optical counterpart.\\ 
\indent A grid of synthetic spectra \citep{mun2005} was then used to match the absorption features to determine approximate spectral parameters of the optical counterpart. We subtracted each template spectrum from the Doppler-corrected average of our Liverpool telescope observations. The grid of spectra included B4V--B7V stars with varying T$_{eff}$ and log $g$. We adopt a $v_{r}$sin$i$ = 45 km s$^{-1}$ from our initial best fit of the line widths to synthetic line profiles. At every point on the synthetic spectra grid, the best fit was found by performing a $\chi^2$ minimization on the residuals of the subtraction. The best fit obtained shows HD314884 as having a T$_{eff}$ = 15,490 $\pm$ 310 K and log $g$ = 3.75 $\pm$ 0.25 dex. From our spectral analysis, the values obtained are in agreement with known B5V--B6V stars contained in Eric Mamajek's online list\footnote[3]{http://www.pas.rochester.edu/$\sim$emamajek/spt/B6V.txt}. We point out that there appear to be no emission lines present in Fig.~1 which argues against mass transfer from the donor star. We confirm this by Doppler correcting and broadening our Liverpool spectra by an appropriate amount found by looking at the FWHM of the absorption lines in the spectral range 3900--5100 {\AA} using IRAF and the SPLOT task. A broadening of 45 km s$^{-1}$ was applied to each spectrum. In a similar manner as above, we then subtracted each from an MK spectral standard B6V template spectrum confirming the absence of residual emission.  

\section{Photometric Period Analysis}
Using the ASAS-3 archival data, we performed an initial period search on the data using the software package PERIOD\footnote[4]{ http://www.starlink.rl.ac.uk/docs/sun167.htx/sun167.html} in the frequency range of 0.01--1000 cycles/day in frequency intervals of 0.00001 cycles/day. PERIOD is a time-series analysis package designed to search for periodicities in data sets. We used the Lomb-Scargle technique to produce a periodogram and search for the maximum peak in the power spectrum for a given data set.\\
\indent The error on the period (along with all of the rest of the errors in this paper, unless otherwise noted) were estimated by using a bootstrapping technique with a Monte Carlo approach much like \citet{tsteeg07}. We copied the target data set 10,000 times where the input dataset is resampled by randomly selecting data points. The data points are allowed to be resampled more than once during the sampling and then each data set `copy' was subjected to fitting an appropriate function to the data and allowing the parameters to vary. For each bootstrap copy, the minimum $\chi^2$ was calculated and the best fit parameters were found. Besides providing the best-fit parameter, the 1$\sigma$ error for each parameter is calculated. This error estimating technique is used for both the photometric and spectroscopic analysis of the data.\\ 
\begin{figure}
	\centering
\includegraphics[width=3.5in,height=3.5in]{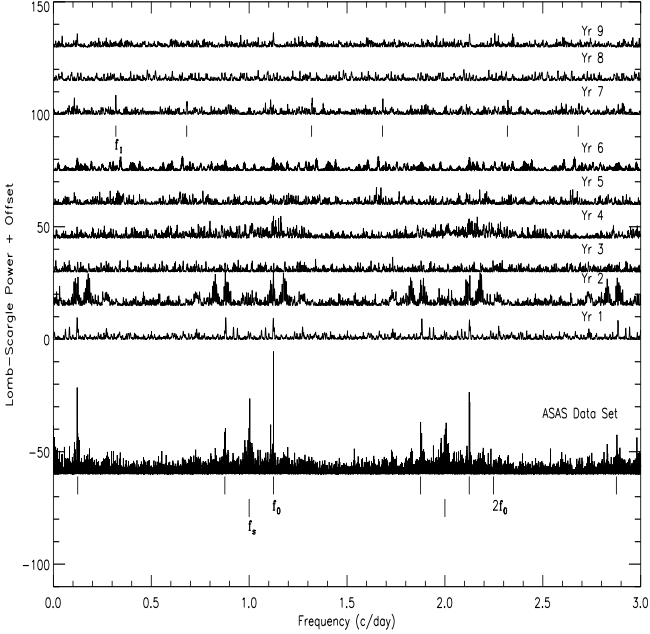}
	\centering
	\caption{Yearly Lomb-Scargle periodograms of HD314884 where 2001 is year 1. The bottom data set is the entire ASAS-3 data set. Several frequency peaks are labeled: $f_0$ corresponds to the 0.889 day period and the aliases; $f_s$ is the one day sampling frequency and aliases; 2$f_0$ is the second harmonic of $f_0$. $f_1$ corresponds to the second distinct period of 3.13 days and aliases. Year 7 shows $f_1$ the best.}
	\centering
	\label{fig:}
\end{figure}  
\indent The ASAS period search confirmed the $V$-band photometric period of 0.889521(12) days which is within the quoted 1$\sigma$ error of the photometric period reported by \citet{hyn12} along with the 8 day alias, the sampling frequencies, and the harmonics (all of which can be seen in Fig.~2). The nine years of ASAS-3 archival data were inspected year by year and as a complete data set to look for any variations on the 0.889521(12) day period. Fig.~2 shows the periodogram of the data. We identify two ``distinct" frequencies when analyzing the data. The 0.889521(12) day period corresponds to a frequency `$f_0$' and will be referred to as such throughout the paper. The second distinct frequency, `$f_1$', corresponds to a 3.1347(56) day period and is found dominating year 7 of the data set in Fig.~2. The aliases of $f_1$ are also labeled in the year 7 data. This signal is present in several other years but weak. The $f_0$ frequency is recovered in years 1, 2, 4, 6, and 9, while no substantial frequency was recovered above the background noise in year 8. In Fig.~2, we label the one day sampling frequency as `$f_s$' along with its harmonics. Years 3 and 5 do not show any significant signal above the 3$\sigma$ level for each of the two data sets.\\
\indent A Lomb-Scargle frequency search of the 0.9 m SMARTS data concluded in a period of 1.517(28) days, corresponding to $f_2$. The light curve is shown in the middle panel of Fig.~3. Neither $f_0$, nor $f_1$ were reproducible in the SMARTS data set and $f_2$ is not at all present above the noise in any of the yearly ASAS-3 data sets in Fig.~2. Folding the SMARTS data on either of these frequencies does very little to convince us they are the correct frequencies for this data set and we conclude that $f_2$ is a {\it third} distinct frequency of this system.\\

\begin{figure}
	\centering
\includegraphics[width=3in]{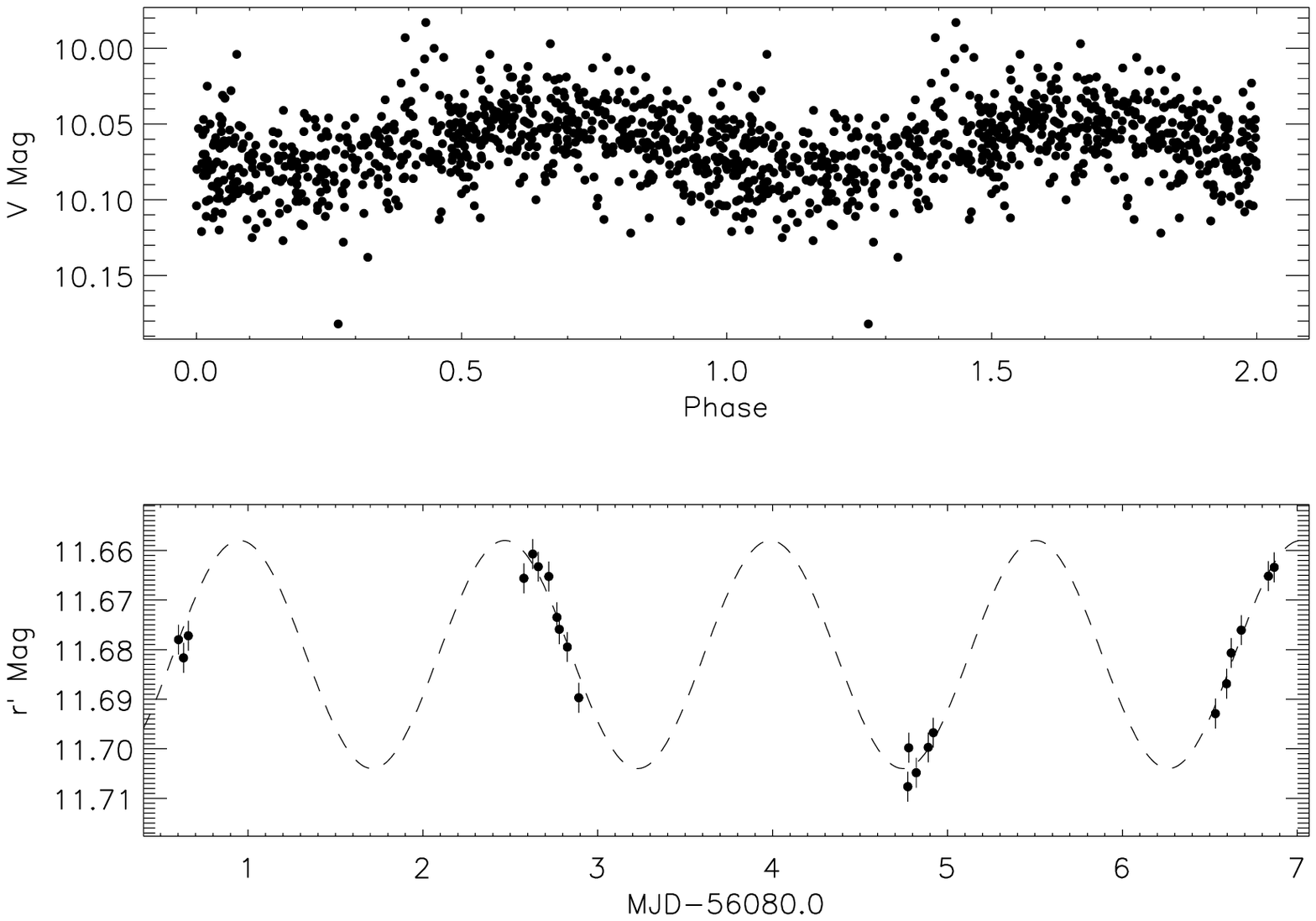}
\includegraphics[width=3.3in,height=1.9in]{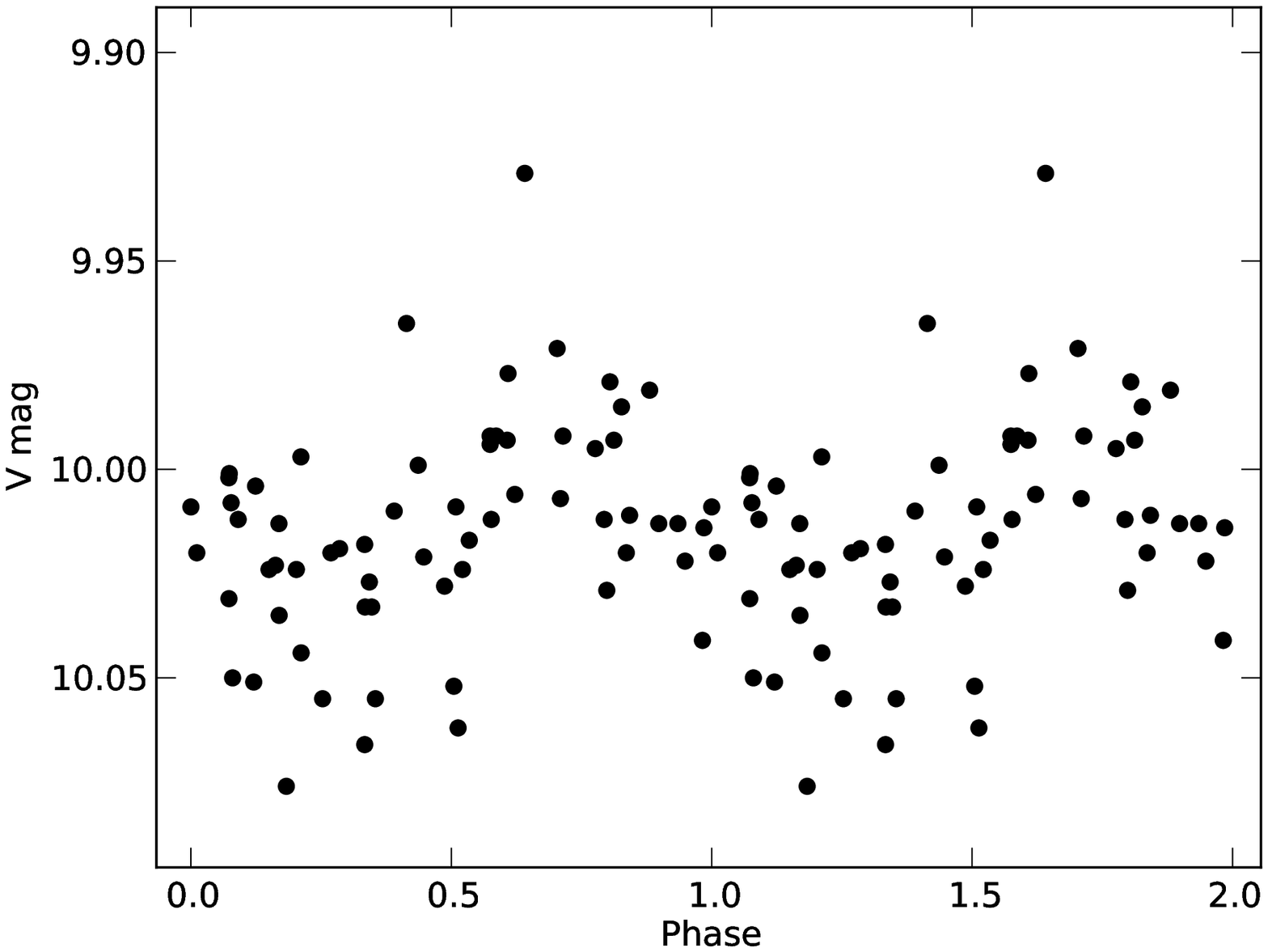}
	\centering
	\caption{The top panel is the phase-folded light curve of the ASAS-3 archival data of HD314884 on the 0.889 day period. The middle panel is the 0.9 m SMARTS light curve with a 1.517(28) day period sine curve plotted over the HD314884 data. The bottom panel shows the phase folded light curve of the ASAS year 7 on the period corresponding to $f_1$ from Fig.~2.}
	\centering
	\label{fig:}
\end{figure}  

\section{Dynamical Analysis}
 \subsection{Radial Velocity Curve}The orbital Doppler shifts of the binary system consisting of HD314884 and its companion were measured from the MgII $\lambda$4481 absorption lines in the spectra. An MK standard B6V template spectrum was chosen and each subsequent target spectrum was cross-correlated against this reference. Using the FXCOR task with other standard IRAF tasks, the radial velocities were determined and a Lomb-Scargle period search was performed on the data. An initial orbital period of P$_{orb}$ = 1.3665(61) days was recovered.\\ 
 \indent The radial velocity curve was then fitted with a sine wave function of the form: y($\phi$) = $\gamma$ + K$_2$ sin(2$\pi \phi$ + $\psi$) with all parameters allowed to vary. The best fit was found by using $\chi^2$ minimization. We find that with a $\chi^2$ = 37.14 and 21 degrees of freedom, this gives a $\chi^2_{red}$ = 1.76 resulting in the following parameters:\\
 P$_{orb}$ = 1.3654 $\pm$ 0.0011 days\\ 
 K$_2$ \hspace*{1mm}      = 49.0 $\pm$ 2.1 km s$^{-1}$\\
 $\gamma$   \hspace*{4mm}= --15.4 $\pm$ 0.6 km s$^{-1}$ \\
 $T_{0}$    \hspace*{2.5mm}= 2456141.395 $\pm$ 0.003 (JD)\\ 
\indent The phase-folded radial velocity curve can be seen in Fig.~4. The errors bars for these parameters are the result of $\Delta\chi^2$ except for P$_{orb}$ which was found from the bootstrapping technique discussed above. The systemic velocity can be prone to systematic calibration errors, but the independent sampling over multiple nights helps mitigate the problem.\\
 \indent We then looked at the full-width at half-maximum variations of several absorption features inherent in the spectra. For the MgII $\lambda$4481 line, we find that there is a peak in the Lomb-scargle periodogram at $f_0$ for the data. The phase-folded light curve of the FWHM of the MgII $\lambda$4481 line can be seen in Fig.~5 with a folded period of 0.889521 days. There is an apparent modulation corresponding to the fundamental frequency of HD314884. 
    
\begin{figure}
	\centering
	\includegraphics[width=3in,height=2.3in]{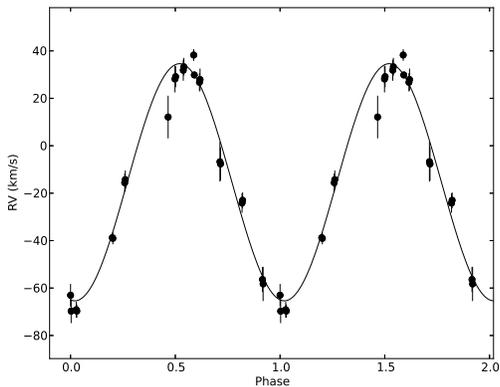}
	\centering
	\caption{The phase-folded radial velocity curve of HD314884 and the unseen companion with the P$_{orb}$ = 1.3654(11) days.}
	\centering
	\label{fig:}
\end{figure}    
 
 \subsection{Masses of the Stellar Components}  The mass function of the unseen companion found from the given parameters is: $f(M_1)$ : M$_1$$ \frac{sin^3i}{(1+q)^2}$ = $\frac{P_{orb}K_2^3}{2\pi G}$ = 0.0157 $\pm$ 0.0021 M$_{\sun}$, where $q$ = $\frac{M_2}{M_1}$ is the mass ratio. The greatest uncertainty comes from the inclination angle, $i$, and the mass of the companion star, M$_1$, when calculating the mass ratio. We used a Monte Carlo method sampling random inclination angles distributed as sin$i$ and assumed B5V--B6V mass range of 4.1 M$_{\sun}$ $\le$ M$_2$ $\le$ 4.5 M$_{\sun}$ (uniformly distributed)$^{3}$. The result is a probability distribution of masses given these prior assumptions showing a peak at $\sim$ 0.8$^{+0.3}_{-0.1}$ M$_{\sun}$ in Fig.~6. The asymmetric errors quoted are due to the fact that the mass distribution is not gaussian. This mass constraint places the unseen companion in the region of an active G-type star or a white dwarf. For the mass distribution data set, we find that the 99\% confidence region encloses a mass range of  0.6--2.3 M$_{\sun}$. Although this mass range does not rule out a neutron star X-ray source, it can rule out a black hole as the X-ray source unless it is of very low mass or the inclination is very low. \cite{ozo10} found that for 16 stellar black holes, the mass distribution peaks at 7.8 $\pm$ 1.2 M$_{\sun}$. The cutoff mass at the low end is $\ge$ 5 M$_{\sun}$ (95\% confidence), indicating a significant lack of black holes in the $\sim$2--5 M$_{\sun}$ range. \cite{farr11} find that a somewhat lower bound of  $\ge$ 4.3 M$_{\sun}$ with 90\% confidence. Their results concerning the low-mass sample are in qualitative agreement with \cite{ozo10}, although the claim is made that their broad model analysis more reliably reveals a better quantitative description of the mass distribution.

\begin{figure}
	\centering
\includegraphics[width=3.0in]{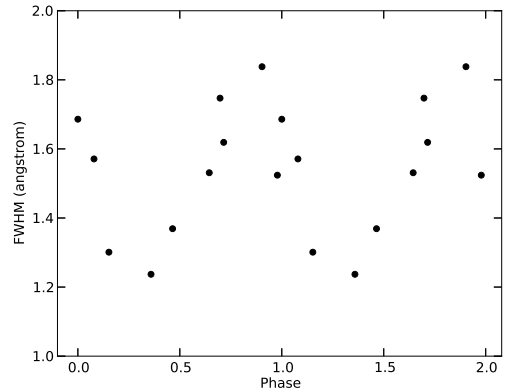}
	\centering
	\caption{The full-width at half-maximum (FWHM) variations of the MgII $\lambda$4481 spectral feature. The data is phase-folded on the 0.889521 day period.}
	\centering
	\label{fig:}
\end{figure}

 \begin{figure}
	\centering
\includegraphics[width=3in]{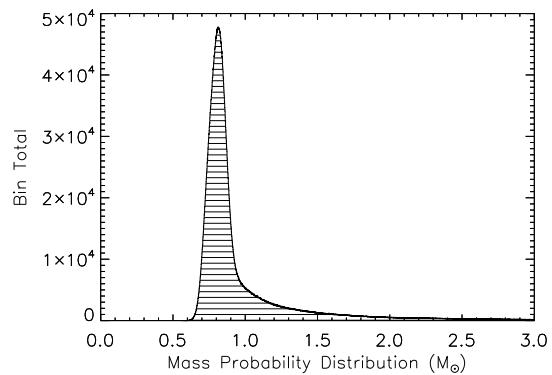}
	\centering
	\caption{The mass distribution of the unseen companion generated by a Monte Carlo sampling routine for random inclination angles distributed as sin$i$ and a B5V--B6V mass range of 4.1 M$_{\sun}$ $\le$ M$_2$ $\le$ 4.5 M$_{\sun}$ distributed uniformly. The peak corresponds to a most probable mass of M$_1$ $\approx$ 0.8 M$_{\sun}$.}
	\centering
	\label{fig:}
\end{figure} 

\section{Discussion} The main observational characteristics of SPBs are a mid-B spectral type and multi-periodic oscillations inherent in their light curves on the order of days. The oscillations can be characterised as either p-mode (frequencies on the order of $\sim$30 minutes) or g-mode (frequencies on the order of $\sim$ days or longer) referring to pressure and gravity modes, respectively. The discovery of HD314884 as an SPB came as a surprise during our analysis of the GBS source CX514. After, the initial photometric period was confirmed using ASAS-3 data, two more distinct periods were uncovered in independent data sets that appear to be non-radial g-mode pulsations. Oscillation frequencies uncovered in photometric studies have frequently been used in precision mode identification to interpret the interior makeup of stars \citep{sae2013} through the use of asteroseismologic techniques. The seismic modeling of SPBs relies on simultaneously fitting the observed frequencies and the empirical values of the $f$-parameter \citep{wal12}. The non-adiabatic complex $f$-parameter is defined as the ratio of the radiative flux change to the radial displacement of the photosphere. With the derived values of $f$ and the observed frequencies, available opacity data can be tested along with the chemical composition and the overshooting efficiency of convective cores \citep{wal12}. The discrepancy between the photometric periods and P$_{orb}$ opens the door for this type of asteroseismic analysis on CX514. To perform accurate and precise mode identification, models that deal with cases where non-linear, tidal interactions are vital must be further developed and refined. \\ 
\indent Given P$_{orb}$ = 1.3654(11) days and a B star mass range of 4.1 M$_{\sun}$ $\le$ M$_2$ $\le$ 4.5 M$_{\sun}$, we expect an orbital separation of 8.8--9.0 R$_{\sun}$ making it one of the shortest known period, single lined binaries containing an SPB, similar to HD24587 \citep{aer99}, but significantly shorter than the SPB sample of \cite{dec00}. With the components being so close, tidal effects can play an important role in exciting pulsation frequencies through deformation of the stellar surface \citep{dec00}. In a binary system with an eccentric orbit, a near resonance occurs  when a stellar frequency is close to a multiple of the orbital frequency \citep{ham11}. Tidal resonance can also occur in close, circular orbits with the addition of g-mode frequencies summing to orbital harmonics. We find that {\bf $f_0$ + $f_1$ $\approx$ 2$f_{orb}$} ( 1.124200(12) day$^{-1}$ + 0.3190(6) day$^{-1}$ = 1.4432(1) day$^{-1}$, where $f_{orb}$ = 0.7323(5) day$^{-1}$ ) indicating a type of higher order, non-linear tidal process \citep{wei12}. None of the identified frequencies found were {\it multiples} of the orbital frequency.\\
\indent The spectral energy distribution (SED) can be seen in Fig.~7. This is constructed with archival data from AAVSO Photometric All-Sky Survey DR7 (APASS) (circles), Tycho (diamonds), DeNIS (triangles), 2MASS (pentagons), and the Galactic Legacy Infrared Mid-Plane Survey Extraordinaire (GLIMPSE) (squares). The model is a Kurucz solar metallicity atmosphere \citep{cas03} with a four way average of T = 15,000 K, log $g$ = 3.5,4.0 and T = 16,000 K, log $g$ = 3.5,4.0. If we assume the spectroscopic effective temperature derived above, we can fit this SED
with an appropriate reddened Kurucz model atmosphere spectrum to estimate a reddening of $E(B-V) \sim0.22$. Assuming a typical
main-sequence absolute $K$ magnitude from Mamajek's compiled list$^3$, the
$K$ band magnitudes from the 2 Micron All Sky Survey (2MASS) and the
Deep Near Infrared Survey of the Southern Sky (DeNIS) imply a distance
of $\sim$1 kpc for HD314884. This suggests that HD314884 has an
L\textsubscript{x} of about 3$\times$10$^{30}$ erg s$^{-1}$.\\
\indent If we assume an active G star as the companion to HD314884, we can
describe the source of soft X-rays as coming from Solar-type coronal
flares. We know from Solar observations that magnetic phenomena are
solely responsible for coronal activity and, in turn, the production
and detection of X-rays. The same holds true for Solar-like G stars
except that stellar flare peak temperatures and emission measures can
be orders of magnitude greater than that of the Sun \citep{joh12}. The
range of L\textsubscript{x} for G stars spans almost 3 orders of
magnitude from a few times 10$^{26}$ erg s$^{-1}$ to
(2--4)$\times$10$^{30}$ erg s$^{-1}$ \citep{mag87}. Our inferred X-ray
luminosity falls near the top end of this range, consistent with a G
star near coronal saturation. The five X-ray photons from HD314884
that were detected by Chandra span the soft end of the spectrum from
0.84--1.81 keV. Such soft X-ray detections have been shown to
originate in plasma temperatures reaching above T$\sim$10$^6$ K
\citep{mag90} in the coronal layer of the star due to flaring
activity.\\
\indent We note the calculated luminosity above would be unusually low if the
X-ray source were a quiescent accreting NS (see Fig.~4 of
\cite{rey11}). \cite{jon07} also point out an exception with 1H
1905+000, which is a LMXB harboring a NS. They inferred an X-ray
luminosity limit of 1.7--2.4$\times$10$^{30}$ erg s$^{-1}$ from 300 ks
of Chandra observations. This is comparable to the luminosity we
infer, so a neutron star companion cannot be ruled out based on the
X-ray luminosity. A binary system with a young pulsar cannot be ruled out either since the inferred X-ray luminosity could support the idea as well.\\
\indent Despite the absence of emission lines in the spectra, we cannot fully rule out mass transfer on to an accretion disc. We do want point out that due to the luminosity of the B6 star, emission in the spectra, if present, may be masked. We doppler corrected the spectra and then subtracted them from one another to look for residual emission lines and found none. Also, if mass transfer were present, we would expect much more pronounced ellipsoidal modulations on the spectroscopic period in either of the data sets much like Fig.~3 in \citet{Rat2013}.\\
\indent We estimate the Roche lobe radius, R$_{L,2}$ for the B5V--B6V from \cite{egg83} to be 4.6--4.8 R$_{\sun}$ using the mass range 4.1 M$_{\sun}$ $\le$ M$_2$ $\le$ 4.5 M$_{\sun}$. The radius of a main sequence B star would be approximately 3.0 R$_{\sun}$ based on the parameters from the Mamajek compilation list (see footnote 3), not yet large enough to start accretion on to the G star by Roche lobe overflow. 

 \begin{figure}
	\centering
\includegraphics[width=3.5in,height=3.5in]{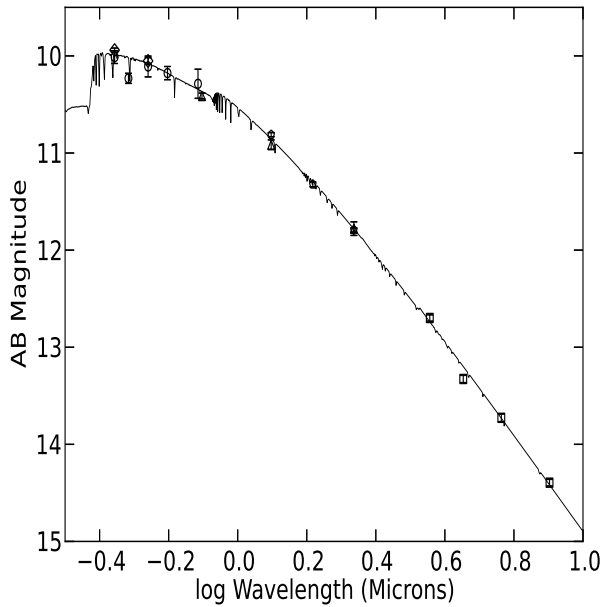}
	\centering
	\caption{The spectral energy distribution for HD314884. The data are drawn from AAVSO Photometric All-Sky Survey DR7 (APASS) (circles), Tycho (diamonds), DeNIS (triangles), 2MASS (pentagons), and the Galactic Legacy Infrared Mid-Plane Survey Extraordinaire (GLIMPSE) (squares).The model is a Kurucz solar metallicity atmosphere with a four way average of T = 15,000 K, log $g$ = 3.5, 4.0 and T = 16,000 K, log $g$ = 3.5, 4.0.}
	\centering
	\label{fig:}
\end{figure} 

\section{Conclusion} We performed a dynamical study of the GBS X-ray source CX514 through optical spectroscopy and photometry. The X-ray source coordinates provided by Chandra align with the bright ($V_{mag}$ = 10.04) B star HD314884, which appears to be the optical counterpart to CX514. The optical spectra are dominated by the absorption lines of HD314884, with no evidence of emission features visible. We have reclassified HD314884 as a B5V--B6V spectral type, slowly pulsating B star with at least 3 distinct oscillation frequencies. An initial photometric period search concluded in detecting two distinct periods in the ASAS-3 archival data. The dominant period of P$_{0}$ = 0.889521(12) days along with a weaker, but prominent, P$_{1}$ = 3.1347(56) days in the ASAS-3 archival data (Fig.~2) supports the claim of an SPB. Upon analyzing the 0.9 m SMARTS data (Fig.~3, middle panel), a third distinct photometric period was found at P$_{2}$ = 1.517(28) days. The observed photometric frequencies are reminiscent of g-mode pulsations and appear to show evidence of non-linear tidal resonances.\\
\indent From phase-resolved spectroscopy, we find an orbital period of 1.3654(11) days from the radial velocity curve with the phase-folded radial velocity curve seen in Fig.~4. After fitting the radial velocity curve with a sinusoidal function and finding the parameters with a $\chi^2$ minimization technique, we solved the mass function $f(M_1)$: M$_1$$ \frac{sin^3i}{(1+q)^2}$ = $\frac{P_{orb}K_2^3}{2\pi G}$ = 0.0157 $\pm$ 0.0021 M$_{\sun}$. Not knowing the inclination angle of the system, a Monte Carlo simulation was used to calculate the mass probability distribution of the unseen companion to HD314884. We obtain a mass of M$_1$ = 0.8$^{+0.3}_{-0.1}$ M$_{\sun}$ for the companion using a uniformly distributed B6 star mass$^3$ for HD314884. We find that this mass constraint places the X-ray source in the region of either an active G star or a white dwarf. Our calculated 99\% confidence region (0.6--2.3 M$_{\sun}$) rules out the possibility of a black hole as the unseen companion emitting X-rays unless it were of very low mass, or of very low inclination, which is unlikely. We cannot fully rule out a neutron star since observed X-ray binaries harboring a NS have been confirmed with masses in this range \citep{kiz13}. Given the derived L\textsubscript{x} of $\sim$3$\times$10$^{30}$ erg s$^{-1}$, and our most probable mass estimate of $\sim$0.8 M$_{\sun}$, we can say that a NS is unlikely for this system. The systemic radial velocity, distance estimate, and the proper motion for HD314884 of --10.7 $\pm$ 1.5 mas yr$^{-1}$ \citep{hog98} is consistent with a massive star that never travels far from the mid plane of the disc. This would rule against a natal kick given to the system if the unseen companion were a NS.

\section*{Acknowledgments}
This work was supported by the National Science Foundation under Grant No. AST-0908789. CBJ acknowledges support from a Graduate Student Research Award (GSRA) administered by the Louisiana Space Grant Consortium (LaSPACE). This research has made use of the SIMBAD database, operated at CDS, Strasbourg, France, and NASA's Astrophysics Data System. Finally we are very grateful to Dr. Eric Mamajek for making available his compilation of stellar colours and temperatures for dwarf stars. The Proposal ID \# for the 2.0 m Liverpool Telescope observing run is JL12B01 with P.I. Thomas J. Maccarone.
Facilities: ASAS, CTIO, LIVERPOOL TELESCOPE, GLIMPSE, DeNIS, 2MASS, TYCHO, APASS DR7.





\end{document}